\documentclass[dvips]{article}
\usepackage{graphicx}
\usepackage{amssymb}
\usepackage{graphicx}
\usepackage{dcolumn}
\usepackage{amsmath}
\usepackage{lscape}
\usepackage{longtable}
\usepackage{subfigure}
\usepackage{color}

\begin{document}
\newcommand{\bd}{\begin{document}}
\newcommand{\ed}{\end{document}}
\newcommand{\bc}{\begin{center}}
\newcommand{\ec}{\end{center}}
\newcommand{\bfr}{\begin{flushright}}
\newcommand{\efr}{\end{flushright}}
\newcommand{\lt}{\left}
\newcommand{\rt}{\right}
\newcommand{\vs}{\vspace}
\newcommand{\hs}{\hspace}
\newcommand{\beq}{\begin{equation}}
\newcommand{\eeq}{\end{equation}}
\newcommand{\lb}{\linebreak}
\newcommand{\pb}{\pagebreak}
\newcommand{\mb}{\makebox}
\newcommand{\fb}{\framebox}
\newcommand{\mc}{\multicolumn}
\newcommand{\ben}{\begin{enumerate}}
\newcommand{\een}{\end{enumerate}}
\newcommand{\bit}{\begin{itemize}}
\newcommand{\eit}{\end{itemize}}
\newcommand{\ol}{\overline}
\newcommand{\un}{\underline}
\newcommand{\lefq}{\lefteqn}
\newcommand{\ba}{\begin{array}}
\newcommand{\ea}{\end{array}}
\newcommand{\beqa}{\begin{eqnarray}}
\newcommand{\eeqa}{\end{eqnarray}}
\newcommand{\beqas}{\begin{eqnarray*}}
\newcommand{\eeqas}{\end{eqnarray*}}
\newcommand{\bfg}{\begin{figure}}
\newcommand{\efg}{\end{figure}}
\newcommand{\bds}{\begin{displaymath}}
\newcommand{\eds}{\end{displaymath}}
\newcommand{\btb}{\begin{tabbing}}
\newcommand{\etb}{\end{tabbing}}
\bc {
\textbf{\huge  Spin and Pseudo-Spin Symmetries in Radial Dirac Equation and Exceptional Hermite Polynomials}  } \ec

\vs{1cm}

\bc
{\it \"Ozlem Ye\c{s}ilta\c{s}$^{*}${\footnote {e-mail : yesiltas@gazi.edu.tr} and Aynur \"{O}zcan$^{*}$  \\
$^{*}$Department of Physics, Faculty of Science, Gazi University,
06500 Ankara, Turkey\\
\vspace{.16cm}

}} \ec \vs{1cm}
\begin{abstract}
\noindent We have generalized the solutions of the radial Dirac equation with a tensor potential under spin and pseudospin symmetry limits to exceptional orthogonal Hermite polynomials family. We have obtained new general rational potential models which are the generalization of the nonlinear isotonic potential families and also energy dependent.
\end{abstract}
\noindent {\bf keyword:}   Dirac equation, spin symmetry, pseudo-spin symmetry, energy dependent potentials. \\

\section{Introduction}
Dirac equation solutions are important for the definition of nuclear shell structure in nuclear physics studies. Assume that an electron  placed in an external electromagnetic field described by a vector potential $A_{\mu}(\vec{x})$. Such a field is placed in the Dirac equation via minimal coupling \cite{d1}. A scalar potential in the Dirac equation is explained as a dependence of the rest mass on position, for instance ,  the potential can be a function of radius about a fixed center which are used to explain  "bag" models of hadrons \cite{d2}. The concept of pseudospin used in shell structure studies was first defined 51 years ago by Hect, K.T, Arima, A. et al \cite{arima}, \cite{adler}.  It has been shown that $\mathcal{SU}(2)$ symmetries of the Dirac equation with scalar and vector potentials in the limit of pseudospin and spin symmetries occuring in the Dirac equation when considering the properties of mesons according to the quark-antiquark model\cite{d3}.  Spin and pseudospin symmetries have been studied for  fifthy years \cite{arima} when pseudospin  doublets were mentioned  in shell-model orbitals with nonrelativistic quantum numbers which are $(n, \ell, j=\ell+1/2)$ and $(n-1, \ell+2, j=\ell+3/2)$ in terms of radial quantum number, orbital and total angular momentum quantum numbers. Both splittings of spin doublets and pseudospin doublets play critical roles in the evolution of magic numbers in exotic nuclei but there were more efforts to understand the origin of those magic numbers. In this way, Hecht and Adler \cite{adler} and Arima, Harvey, and Shimizu \cite{arima} found a degeneracy between two single-particle states with quantum numbers $(n, \ell, j=\ell+1/2)$ and $(n-1, \ell+2, j=\ell+3/2)$. The experimental analysis of spin and pseudospin symmetries can be found in \cite{gino}. Moreover, the relevance of QCD sum rules with pseudospin symmetries were shown in \cite{gino2}. Spin symmetry shows up in the spectrum of mesons with one heavy quark, also hadrons, antinucleons have spin symmetry \cite{gino2}.  Pseudospin symmetry in heavy nuclei  originates from the nucleons in a nucleus move in an attractive scalar and repulsive vector potential \cite{pss1}, \cite{pss2} which is used to explain the observed degeneracies of some shell-model orbitals in nuclei. Pseudospin symmetry is  an approximation
 in a real nuclei and it is shown that the pseudospin symmetry is examined in the framework of relativistic Hartree-
Bogoliubov theory for some examples \cite{meng}.  Furthermore, the energy dependency in potential functions in the Hamiltonian operators have been attracted much attention \cite{ED1}, \cite{ED2}, \cite{ED3}.  Among the physical applications, one can find  models of heavy quark systems \cite{heavy}, semiconductors \cite{semi} and energy loss characteristics of an electron probe \cite{probe}. Accordingly, spin symmetry and pseudospin symmetry were investigated  for the systems whose potential functions are energy dependent \cite{4}, \cite{5}, \cite{66}. Energy dependent Hamiltonians first mentioned in 1927 through Pauli  Schrodinger equation \cite{pauli}. They play a role both in relativistic and non-relativistic quantum mechanics. For instance, some types of  potentials with power-law radial shapes in the three dimensions \cite{yekken}, exact solutions of the Dirac equation which has spin and pseudospin symmetries \cite{hassan}, asymptotic iteration methods approach to energy dependent Hamiltonians \cite{barakat}. This concept, which is suitable for spherical nuclei, has been proven to be valid for deformed nuclei \cite{raju} and has been successfully applied to explain many phenomena in nuclear structural physics studies \cite{6,7,8,9,10,11,12}.

 Moreover, analytical  solutions of the Dirac equation   under the spin and the pseudo-spin symmetry limits in the cosmic string spacetime \cite{lima}, scattering problem of the Dirac equation produced by the Huth\'{e}n potential in a cosmic string background \cite{hossein} are recent studies in this context. Recently there has been a growing interest in exceptional orthogonal polynomials. We can note that  the classical orthogonal polynomials are the only orthogonal polynomials which are solutions of a differential equation
of Sturm-Liouville type. Exceptional orthogonal polynomials are complete orthogonal polynomial systems with respect to a positive measure extending the classical families of Hermite, Laguerre and Jacobi. The exceptional orthogonal polynomials are not included by the Bochner’s classification theorem. They have first discovered in 2009 \cite{kamran} and been studied in quantum mechanics by the authors \cite{sasaki}, \cite{quesne}. Recently, these exceptional orthogonal polynomials were generated by means of the Darboux-Crum transformation \cite{darboux}, physical models studied in quantum mechanics which  involve  Fokker-Planck equations and exceptional Laguerre polynomials \cite{ho}.
In the recent paper, we are interested in models that admit solutions in terms of exceptional orthogonal polynomial systems such as exceptional Hermite polynomials. We note that one can search the rationally extended potential models through the recent studies on spin and pseudspin symmetries. Especially, we start with the radial Dirac equation with unknown the scalar, vector and tensor potential forms. We use the relevant point transformations to arrive at  exceptional Hermite polynomials which are convenient for our system.

This paper is organized as follows: Section II includes the radial Dirac equation within tensor potential and spin and pseudospin symmetry cases are given. In Section III, we give the necessary transformations for Dirac equations and exceptional Hermite polynomials are introduced. At the same time, we give the spin and pseudospin symmetric solutions. The demonstrations of the probability density and physical potential are given for each case. We conclude our results in Section IV.

\section{Dirac Equation and Spin/Pseudo-Spin Symmetries}
For a particle interacting with an external electric field through the magnetic moment is expressed by a Dirac Hamiltonian with a tensor potential \cite{gino}.

 Let us write the Dirac equation in the form given in the References   \cite{1}, \cite{akcay}
\begin{eqnarray}\label{01}
% \nonumber to remove numbering (before each equation)
  (\frac{d}{dr}+\frac{k}{r}-U(r))F(r) &=& (M+E-V(r)+S(r))G(r) \\
 (\frac{d}{dr}-\frac{k}{r}+U(r))G(r) &=& (M-E+V(r)+S(r))F(r) \label{02}
\end{eqnarray}
where $S(r)$ and $V(r)$ are scalar and vector potentials respectively. Here $U(r)$ is the tensor potential term. Then, we can continue to study (\ref{01}) and (\ref{02}) within spin and pseudospin symmetries.
\\
\emph{\textbf{Case 1: spin symmetry:}}
\\
For the spin symmetric case we take $S(r)=V(r)$ in (\ref{01}) and (\ref{02}), and obtain the couple of second order linear differential equations
\begin{equation}\label{0010}
  -\frac{d^{2}F}{dr^{2}}+\left(2(E+M)V(r)+U(r)^{2}-\frac{2kU(r)}{r}+\frac{k(k+1)}{r^{2}}+U'(r)\right)F(r)=(E^{2}-M^{2})F(r)
\end{equation}
and
\begin{equation}\label{2}
\begin{split}
  -\frac{d^{2}G(r)}{dr^{2}}&-\frac{2V'(r)}{E-M-2V(r)}\frac{dG(r)}{dr}\\+& \left(\frac{k(k-1)}{r^{2}}-\frac{2kU(r)}{r}+U(r)^{2}+2(E+M)V(r)-U'(r)
  -  \frac{2(-k+rU(r))V'(r)}{r(E-M-2V(r))}\right)G(r)= (E^{2}-M^{2})G(r).
  \end{split}
\end{equation}
\\
\emph{\textbf{Case 2:pseudo-spin symmetry:}}
\\
If we take $V(r)=-S(r)$, then, our system (\ref{01})-(\ref{02}) turns into
\begin{equation}\label{1s}
  -\frac{d^{2}G}{dr^{2}}+\left(2(E-M)V(r)+U(r)^{2}-\frac{2kU(r)}{r}+\frac{k(k-1)}{r^{2}}-U'(r)\right)G(r)=(E^{2}-M^{2})G(r)
\end{equation}
\begin{equation}\label{2s}
\begin{split}
  -\frac{d^{2}F(r)}{dr^{2}}&-\frac{2V'(r)}{E+M-2V(r)}\frac{dF(r)}{dr}\\+& \left(\frac{k(k+1)}{r^{2}}-\frac{2kU(r)}{r}+U(r)^{2}+2(E-M)V(r)+U'(r)
  +\frac{2(-k+rU(r))V'(r)}{r(E+M-2V(r))}\right)F(r)=(E^{2}-M^{2})F(r).
  \end{split}
\end{equation}
\section{Transformations and Solutions}
If we introduce a second order differential equation below,
\begin{equation}\label{03}
  G''(r)+\bar{g}(r)G'(r)+\bar{h}(r)G(r)=0,
\end{equation}
where
\begin{equation}\label{003}
  \bar{g}(r)=\frac{2V'(r)}{E-M-2V(r)}
\end{equation}
and
\begin{equation}\label{4}
  \bar{h}(r)=-\left(\frac{k(k-1)}{r^{2}}-\frac{2kU(r)}{r}+U(r)^{2}+2(E+M)V(r)-U'(r)
  -  \frac{2(-k+rU(r))V'(r)}{r(E-M-2V(r))}-(E^{2}-M^{2})\right),
\end{equation}
which are coefficients of first order derivative and derivative-free terms in (\ref{03}). Let us apply a mapping to (\ref{03}) given below
\begin{equation}\label{004}
  G(r)=\exp(\int^{r} \mu(t)dt)\bar{G}_1(r),
\end{equation}
then, we obtain
\begin{equation}\label{0030}
  \bar{G}_{1}''(r)+g(r)\bar{G}_{1}'(r)+h(r)\bar{G}_{1}(r)=0,
\end{equation}
where
\begin{eqnarray}\label{g1}
% \nonumber to remove numbering (before each equation)
  g(r) &=& 2\mu(r)+\frac{2V'(r)}{E-M-2V(r)} \\
  h(r) &=&  -(\frac{k(k-1)}{r^{2}}-\frac{2kU(r)}{r}+U(r)^{2}+2(E+M)V(r)-U'(r) \label{h1}
  -  \frac{2(-k+rU(r))V'(r)}{r(E-M-2V(r))}-\\ \nonumber
  &  & \mu(r)^{2}-\frac{2\mu(r)V'(r)}{E-M-2V(r)}-\mu'(r)). \nonumber
\end{eqnarray}
Now, (\ref{0030}) can be discussed within a point transformation given below:
\begin{equation}\label{5}
  \bar{G}_{1}(r)=\frac{\exp\left(\int^{^{r}}\frac{g(r(z))r'(z)}{2}dz\right)}{\sqrt{r'(z)}}\bar{G}_{1}(z),
\end{equation}
then, (\ref{0030}) turns into
\begin{equation}\label{60}
  \bar{G}''_{1}(r)+\left(-\frac{g(r(z))^{2}r'(z)^{2}}{4}+h(r(z))r'(z)^{2}+\frac{g(r(z))r'(z)^{2}}{2}-\frac{g'(r(z))r'(z)^{2}}{2}-
  \frac{3(r''(z))^{2}}{4r'(z)^{2}}+\frac{r'''(z)}{2r'(z)}\right)\bar{G}_{1}(r)=0.
\end{equation}
Our goal is to transform the systems (\ref{2}) and (\ref{2s}), which are equations of spin and pseudospin symmetric cases, into (\ref{60}).
\subsection{Exceptional Hermite Polynomials}
 A partition of length of $t$ is a finite decreasing sequence of positive integers $\lambda=(\lambda_1 \geq \lambda_2...\geq \lambda_t \geq 1)$ \cite{kamran}. The generalized Hermite polynomials are given by the Wronskian determinant
 \begin{equation}\label{7}
   H_{\lambda}= Wr[H_{\lambda_{t}},...H_{_{\lambda_{2}+t-2}, H_{\lambda_{1}+t-1}}].
 \end{equation}
If $\lambda_{2k-1}=\lambda_{2k}, k=1,2,...r/2$, then, $\lambda$ is an even partition, $H_{\lambda}$ has no zeros:
\begin{equation}\label{8}
  W_{\lambda}(r)=\frac{e^{-r^{2}}}{H_{\lambda}(r)^{2}},~~r\in (-\infty, \infty),
\end{equation}
where $W_{\lambda}(r)$ is the weight function.  If we use a polynomial of degree $n$ is $P_{n}$,
\begin{equation}\label{9}
  P_{n}=Wr[H_{\lambda_{t}}, H_{\lambda_{t-1}+1},...H_{\lambda_{2}+t-2}, H_{\lambda_{1}+t-1, H_{n-\mid\lambda \mid}+t}],
\end{equation}
$P_n$ satisfies the differential equation below:
\begin{equation}\label{10}
  P''_{n}(r)-2(r+\frac{H'_{\lambda}}{H_{\lambda}})P'_{n}(r)+(\frac{H''_{\lambda}}{H_\lambda}+2r\frac{H'_\lambda}{H_{\lambda}}+2n-2\mid\lambda\mid)P_{n}(r)=0.
\end{equation}
In addition, the weighted product and orthogonality aspects are given respectively,
\begin{eqnarray}
% \nonumber to remove numbering (before each equation)
  P_n(r)P_m(r) W_\lambda(r) &=& \frac{d}{dr}(\frac{P_n(r)P'_{m}(r)-P'_{n}(r)P_{m}(r)}{2(n-m)}) \\
  \int^{\infty}_{-\infty} P_n(r)P_m(r) W_\lambda(r)dr &=&0, ~~n,m \in \mathbb{N}.
\end{eqnarray}
\subsection{Solutions}
\emph{\textbf{Spin Symmetric Case Solutions}:}\\
If we compare (\ref{10}) and (\ref{0030}), we understand that
\begin{eqnarray} \label{11}
% \nonumber to remove numbering (before each equation)
  g(r) &=& -2(r+\frac{H'_{\lambda}(r)}{H_{\lambda}(r)}) \\
  h(r) &=& \frac{H''_{\lambda}(r)}{H_\lambda(r)}+2r\frac{H'_\lambda(r)}{H_{\lambda}(r)}+2n-2m. \label{12}
\end{eqnarray}
Now we can plug (\ref{11}) and (\ref{12}) in (\ref{60}). We obtain,
\begin{equation}\label{13}
  \bar{G}''_{1}(r(z))+\left((1-2m+2n-r(z)^{^{2}})r'(z)^{2}+\left(-\frac{2P'_{\lambda}(r(z))^{2}}{P_{\lambda}(r(z))^{2}}+
  \frac{2P''_{\lambda}(r(z))}{P_{\lambda}(r(z))}\right)(r'(z)^{2})-\frac{3(r''(z)^{2})}{4(r'(z)^{2})}+\frac{r'''(z)}{2r'(z)}\right)\bar{G}_{1}(r(z))=0.
\end{equation}
Here, we choose the partition and $r(z)$ as below:
\begin{equation}\label{14}
  \lambda=\{1,1\}, ~~r(z)=\alpha z.
\end{equation}
And we note that the Wronskian of the Hermite polynomials can be found as
\begin{equation}\label{15}
  P_{\lambda}(r)=H_{1,1}(r)=Wr[H_1,H_2](r)=4(1+2r^{2}).
\end{equation}
Then, the solutions of (\ref{13}) become
\begin{equation}\label{155}
  \bar{G}_{1}(z)=N_{n} e^{-\frac{r(z)^{2}}{2}}\frac{1}{H_{1,1}(r(z))}\frac{1}{\sqrt{r'(z)}}H_{\lambda,n}(r(z)).
\end{equation}
Using (\ref{14}) and (\ref{15}), we may re-write (\ref{13}) as
\begin{equation}\label{16}
  \bar{G}''_{1}(z)+\left(2\alpha^{2}(n-m)-\alpha^{4}z^{2}-\frac{32\alpha^{2}z^{2}}{(1+2\alpha^{2}z^{2})^{2}}+
  \frac{8\alpha^{2}}{1+2\alpha^{2}z^{2}}\right)\bar{G}''_{1}(z)=0.
\end{equation}
It can be seen that the eigenvalues in (\ref{16}) are related to $2\alpha^{2}(n-m)$ where $m$ may take any integer value. Thus, the eigenvalues for (\ref{0010}) and (\ref{2}) are given by
\begin{equation}\label{17}
  E_{n}=\pm \sqrt{M^{2}+2\alpha^{2}(n-m)}.
\end{equation}
The scalar potential function in (\ref{16}) is known as nonlinear isotonic oscillator in the literature \cite{iso1}, \cite{iso2}, \cite{iso3}, \cite{iso4}. Hence, we can calculate the whole solutions within the polynomial $H_{\lambda,n}(r(z)$:
\begin{equation}\label{170}
 G_{n}(z)=N_{n}e^{-\frac{\alpha^{2}z^{2}}{2}}\frac{1}{4\sqrt{\alpha}(1+2\alpha^{2}z^{2})}\left(n(n-1)(2\alpha^{2}z^{2}+n+1)H_{n-2}(\alpha z))-(n+1)\left(\frac{2n\alpha z}{\sqrt{n+1}}H_{n-1}(\alpha z)-H_{n}(\alpha z)\right)\right).
\end{equation}

We can get $\mu(r)$, $V(r)$ and $U(r)$ using (\ref{11}),  (\ref{12}), (\ref{g1}) and (\ref{h1}):
\begin{equation}\label{mu}
  \mu(r)=\frac{1}{2}\left(-2(r+\frac{4r}{1+2r^{2}})-\frac{2V'(r)}{E-M-2V(r)}\right),
\end{equation}
\begin{equation}\label{18}
  V(r)=\frac{-9+13r^{2}+4r^{6}}{2(E+M)(1+2r^{2})^{2}},
\end{equation}
and
\begin{equation}\label{19}
  U(r)=\frac{k}{r}+\frac{V'(r)}{E-M-2V(r)}.
\end{equation}
Then, we can turen back to (\ref{01}) which can turn into
\begin{equation}\label{20}
\begin{split}
   -\frac{d^{2}F}{dr^{2}}+(-1+r^{2}+
   \frac{8(-1+2r^{2})}{(1+2r^{2})^{2}} + \frac{3(C_1+C_2r^{2}+C3r^{4})}{(-9+13r^{2}+4r^{6}-(E^{2}-M^{2})(1+2r^{2}))^{2}}&+\\
   \frac{2(-83+6(E^{4}+M^{4})+10r^{2}+4r^{4}-6M^{2}(r^{2}+3)+6E^{2}(3-2M^{2}+r^{2}))}{(-9+13r^{2}+4r^{6}-(E^{2}-M^{2})(1+2r^{2})^{2})^{2}})F(r)=(E^{2}-M^{2})F(r)
\end{split}
\end{equation}
where
\begin{equation}\label{C1}
  C_1=(9+E^{2}-M^{2})(-39+4E^{4}+(12-8M^{2})E^{2}+4M^{2}(M^{2}-3)),
\end{equation}
\begin{equation}\label{C2}
  C_2=4(169+4E^{6}+131M^{2}-12E^{4}M^{2}-4M^{6}+E^{2}(-131+12M^{4})),
\end{equation}
\begin{equation}\label{C3}
  C_3=4(81+4E^{6}+43M^{2}+16M^{4}-4M^{6}+4E^{4}(4-3M^{2})+E^{2}(-43-32M^{2}+12M^{4})).
\end{equation}
The solutions $F(r)$ should satisfy the equations below for the spin-symmetric case $S(r)=V(r)$:
\begin{eqnarray}
% \nonumber to remove numbering (before each equation)
  F'(r)+(\frac{k}{r}-U(r))F(r) &=& (M+E)G(r)  \\
  G'(r)+(-\frac{k}{r}+U(r))G(r) &=& (M-E+2V(r))F(r).
\end{eqnarray}
Thus, we can easily get the solutions of $F(r)$ using $G(r)$ solutions in this system. We can't give the solutions $F(r)$ here because of lack of simplicity. We show the probability density function graph changing with position for  the different principal quantum numbers as below.
\begin{figure}[h!]
    \centering
    \includegraphics[scale=0.8]{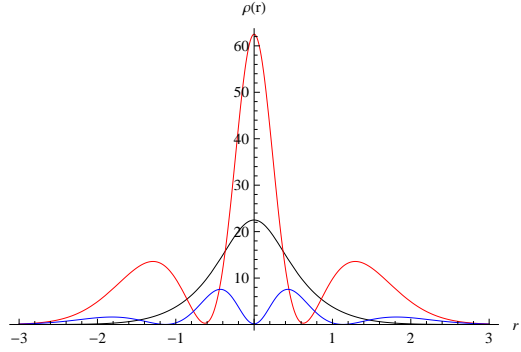}
    \caption{Probability density  $\rho(r)=\| F_{n} \|^{2}+ \| G_{n} \|^{2} $ versus $r$. For the black curve $n=2$, red curve $n=4$ and blue curve $n=5$.}
    \label{fig:my_label}
\end{figure}
\begin{figure}[h!]
    \centering
    \includegraphics[scale=0.8]{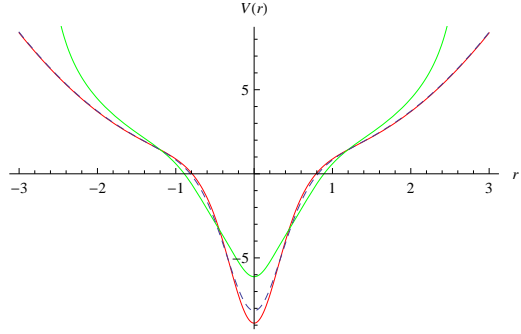}
    \caption{Potential graph for the potential function given in (\ref{0010}). $n=2$, $n=5$ and $n=8$ for the red, green and dashed curves respectively. }
    \label{fig:my_label}
\end{figure}
\begin{figure}[h!]
    \centering
    \includegraphics[scale=0.8]{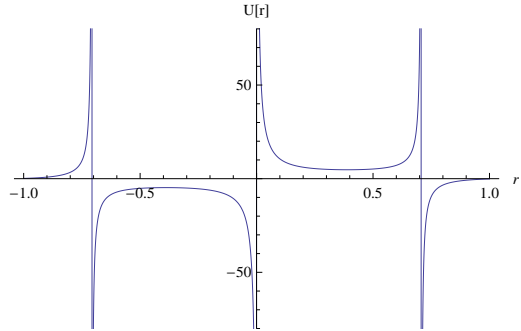}
    \caption{Potential graph for the tensor potential function found in (\ref{19}), $k = 1, \alpha = 0.10, M = 1, n = 2$.}
    \label{fig:my_label}
\end{figure}
\\
\emph{\textbf{Pseudospin Symmetric Case Solutions}:}\\
We can begin with (\ref{2s}) which looks appropriate for our transformation mentioned above:
\begin{equation}\label{21}
   F''(r)+\bar{g}(r)F'(r)+\bar{h}(r)F(r)=0,
\end{equation}
where
\begin{equation}\label{22}
\bar{g}(r)=\frac{2V'(r)}{E+M-2V(r)}
\end{equation}
and
\begin{equation}\label{23}
  \bar{h}(r)=-\left(\frac{k(k+1)}{r^{2}}-\frac{2kU(r)}{r}+U(r)^{2}+2(E-M)V(r)+U'(r)
  +\frac{2(-k+rU(r))V'(r)}{r(E+M-2V(r))}\right)
\end{equation}
\begin{equation}\label{24}
  F(r)=\exp(\int^{r} \mu(t)dt)\bar{F}_1(r).
\end{equation}
Using (\ref{24}) in (\ref{2s}), we obtain
\begin{equation}\label{030}
  \bar{F}_{1}''(r)+g(r)\bar{F}_{1}'(r)+h(r)\bar{F}_{1}(r)=0,
\end{equation}
where
\begin{eqnarray}\nonumber
% \nonumber to remove numbering (before each equation)
  g(r) &=& 2\mu(r)+\frac{2V'(r)}{E+M-2V(r)} \\
  h(r) &=& -\left(\frac{k(k+1)}{r^{2}}-\frac{2kU(r)}{r}+U(r)^{2}+2(E-M)V(r)+U'(r)
  +\frac{2(-k+rU(r))V'(r)}{r(E+M-2V(r))}-\mu(r)^{2}-\mu'(r)\right). \nonumber
\end{eqnarray}
Following the same procedure we can transform (\ref{030}). Thus we have,
\begin{equation}\label{5}
  \bar{F}_{1}(r)=\frac{\exp\left(\int^{^{r}}\frac{g(r(z))r'(z)}{2}dz\right)}{\sqrt{r'(z)}}\bar{F}_{1}(z),
\end{equation}
then, (\ref{2s}) turns into
\begin{equation}\label{6}
  \bar{F}''_{1}(z)+\left(-\frac{g(r(z))^{2}r'(z)^{2}}{4}+h(r(z))r'(z)^{2}+\frac{g(r(z))r'(z)^{2}}{2}-\frac{g'(r(z))r'(z)^{2}}{2}-
  \frac{3(r''(z))^{2}}{4r'(z)^{2}}+\frac{r'''(z)}{2r'(z)}\right)\bar{F}_{1}(z)=0.
\end{equation}
In this case, following the similar procedure, one can obtain the functions $\mu(r), V(r)$ and $U(r)$ as
\begin{eqnarray}
% \nonumber to remove numbering (before each equation)
  \mu(r) &=& -(r+\frac{4r}{1+2r^{2}})-\frac{V'(r)}{E+M-2V(r)} \\
  V(r) &=& \frac{-9+13r^{2}+4r^{6}}{2(E-M)(1+2r^{2})} \\ \label{V2}
  U(r) &=& \frac{k}{r}+\frac{r(49-26r^{2}+12r^{4}+8r^{6})}{(1+2r^{2})(-9+13r^{2}+4r^{6}-(E^{2}-M^{2})(1+2r^{2})^{2})}. \label{U2}
\end{eqnarray}
Hence, $F(r)$ solutions can be introduced as
\begin{equation}\label{F}
  F_{n}(r)=N_{n}N_{n}e^{-\frac{r^{2}}{2}}\frac{1}{4\sqrt{\alpha}(1+2r^{2})}\left(n(n-1)(2r^{2}+n+1)H_{n-2}(r))-(n+1)\left(\frac{2nr}{\sqrt{n+1}}H_{n-1}(r)
  -H_{n}(r)\right)\right).
\end{equation}

\begin{figure}[h!]
    \centering
    \includegraphics[scale=0.8]{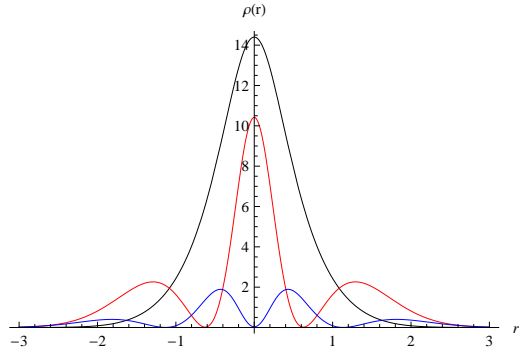}
    \caption{Probability density $\rho(r)=\| F_{n} \|^{2}+ \| G_{n} \|^{2} $ versus $r$ graph of the solutions (\ref{1s}) and (\ref{2s}). For the black curve $n=2$, red curve $n=4$ and blue curve $n=5$.}
    \label{fig:my_label}
\end{figure}
\begin{figure}[h!]
    \centering
    \includegraphics[scale=0.8]{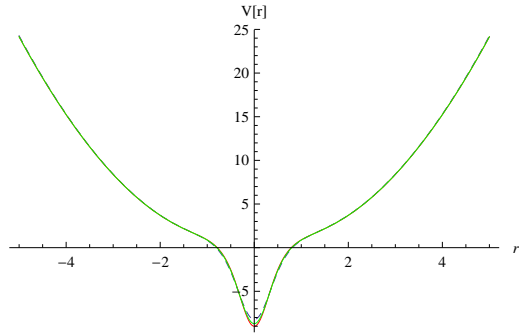}
    \caption{The potential function of (\ref{1s}).  $\alpha=10$, $n=2$ for the dashed curve,  $\alpha=50, n=4$ red curve, $\alpha=10, n=4$ green curve.}
    \label{fig:my_label}
\end{figure}
\begin{figure}[h!]
    \centering
    \includegraphics[scale=0.8]{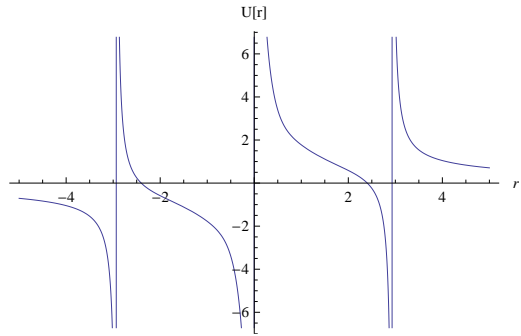}
    \caption{The tensor potential function of (\ref{U2}), $M = 1$, $\alpha = 2, n = 2, k = 2$.}
    \label{fig:my_label}
\end{figure}

%%%%%%%%%%%%%%%%%%%%%%%%%%%%%%%%%%%%%%%%%%%%%%%%%%%%%%%%%%%%%%%%%%%%%%%%%%%%%%%%%%%%%%%%%%%%%%%%%%%%%%%%%%%%%%%%%%%%%%%%%%%%%%%%%%%%%%%%%%%%%%%%%%%%%%%%%%%%
%%%%%%%%%%%%%%%%%%%%%%%%%%%%%%%%%%%%%%%%%%%%%%%%%%%%%%%%%%%%%%%%%%%%%%%%%%%%%%%%%%%%%%%%%%%%%%%%%%%%%%%%%%%%%%%%%%%%%%%%%%%%%%%%%%%%%%%%%%%%%%%%%%%%%%%%%%%%%
\newpage
\section{Conclusions}
There has been a  growing research interest in nonlinear isotonic oscillator models for ten years. This potential model can be defined as  nonpolynomial which corresponds to an oscillator plus a rational term \cite{iso1}. Among the research of potential models for the quark-antiquark systems which are not including such nonlinear isotonic models, the recent work may be an example. In this work, we have used exceptional Hermite polynomial solutions for both spin and pseudospin symmetric cases in the radial Dirac equation to derive  the corresponding tensor, scalar and vector potentials. Thus, we have followed a different path from most previous studies as extending rational potential models to the spin and pseudospin symmetric Dirac equations. Quite importantly, we have observed that the generated potentials are obatined as energy dependent in both cases. The energy eigenvalues of both systems are common and corresponding solutions are given in terms of exceptional Hermite poynomials. We have given the plots of potentials and conventional probability functions(Figures $1$ and $4$) for each case. In the spin symmetric case, the depth of the potential well which is shown in Figure $2$ is increasing with respect to the quantum number $n$.In Figure $5$, one may see that the potential depth is very slowly increasing for greater values of  $\alpha$ and $n$. In the Figures $3 $ and $6$, tensor potentials can be seen for spin and pseudospin symmetric states. In both cases, tensor potential is repulsive.

%%%%%%%%%%%%%%%%%%%%%%%%%%%%%%%%%%%%%%%%%%%%%%%%%%%%%%%%%%%%%%%%%%%%%%%%%%%%%%%%%%%%%%%%%%%%%%%%%%%%%%%%%%%%%%%%%%%%%%%%%%%%%%%%%%%%%%%%%%%%%%%%5
%%%%%%%%%%%%%%%%%%%%%%%%%%%%%%%%%%%%%%%%%%%%%%%%%%%%%%%%%%%%%%%%%%%%%%%%%%%%%%%%%%%%%%%%%%%%%%%%%%%%%%%%%%%%%%%%%%%%%%%%%%%%%%%%%%%%%%%%%%%%%%%%%%%%%%

\end{document}